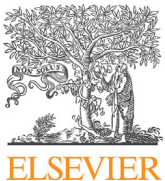
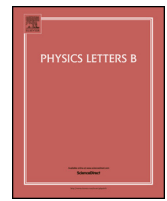

# Multipole expansion for the electron-nucleus scattering at high energies in the unified electroweak theory

Z.P. Luong [a], M.T. Vo [b],*

[a] *Van Lang University, Ho Chi Minh City, Viet Nam*
[b] *Mien Tay Construction University, Vinh Long City, Viet Nam*

A B S T R A C T

The paper presents the multipole expansion for the electron-nucleus scattering cross section at high energies within the framework of the unified electroweak theory. The electroweak currents of the nucleus are expanded into the simple components with definite angular momenta, called the multipole form factors. The multipole expansion of the cross section is a consequence of the above expansion. Besides the familiar electromagnetic form factors, there are also the vector and axial form factors, respectively, related to weak interactions. To determine multipole form factors, general formulas for the calculation of reduced matrix elements are established using the fractional parentage coefficient method and the multiparticle shell model. Calculation of them enables us to obtain more detailed information about the nuclear structure and elucidate the role played by the weak interaction in the high-energy reaction mechanisms.

## 1. Introduction

The method of studying the nuclear structure by electron scattering, or more generally, lepton scattering, is highly effective because it provides many detailed results about the inner structure of the nuclei, especially when the attained electron energy becomes much higher.

In order to relate the structure of the nucleus with the scattered electrons, the best way is to expand the scattering amplitude into multipole components, each term corresponds to a definite angular momentum $Lm$. Weigert-Rose [1] and Willey [2] were the first to perform a complete expansion when the interaction is purely electromagnetic, and the expansion was improved afterward by Donnelly-Raskin [3]. Owing to this expansion, every partial amplitude corresponding to each multipole can be calculated in detail for nuclei, and it clarifies many properties of the nuclei.

At high energies that the electron accelerators reached (about 105 GeV at CERN since 2000), the electron-nucleus scattering must be described by the unified electroweak theory. The neutral current effects in nuclei have also been studied in detail by Donnelly-Peccei [4] within the above-mentioned theory. We shall extend the Weigert-Rose method to expand the scattering amplitude in this case. For simplicity, we consider only the scattering in which the nuclei are unoriented, i.e. the nuclear spins have isotropic distribution. Next, we infer general expressions of multipole operators within the multiparticle shell model and then give formulas for the calculation of multipole form factors without any approximations (for example, impulse approximation [5]). Finally, specific calculations are carried out for the spin-3/2 nucleus.

## 2. Multipole expansion for the scattering cross section

We shall consider the scattering of electrons at high energies, of order GeV, on the nuclei. In order to apply the Born approximation, the target nuclei are supposed to be light such that they satisfy the inequality $Z\alpha < 1$, $\alpha$ is the electromagnetic coupling constant. At this energy scale, in the scattering the electron exchanges a photon $\gamma$ and an intermediate neutral boson $Z^0$ with the nucleus, the scattering amplitude is of the form:

---

\* Corresponding author.
  *E-mail addresses:* lzphu44@gmail.com (Z.P. Luong), vominhtruong@mtu.edu.vn (M.T. Vo).

https://doi.org/10.1016/j.physletb.2023.138095
0370-2693/© 2023 The Author(s). Published by Elsevier B.V. This is an open access article under the CC BY license (http://creativecommons.org/licenses/by/4.0/). Funded by SCOAP$^3$.

$$M_{fi} = -\frac{4\pi}{Q^2}\left[\alpha \bar{u}'\gamma_\mu u J_F^\mu(Q) + \lambda \bar{u}'\gamma_\mu(g_V + g_A\gamma_5)u J_Z^\mu(Q)\right], \quad (1)$$

where $u = u(K, S)$ and $u' = u(K', S')$ are the (spinor) state amplitudes of the electron before and after scattering respectively, $Q = K - K' = (\omega, \mathbf{q})$ is the 4-momentum transfer, $K = (\varepsilon, \mathbf{k})$ and $K' = (\varepsilon', \mathbf{k}')$ are the 4-momenta, $S$ and $S'$ are the 4-spins, $J_F^\mu(Q)$ is the electromagnetic current and $J_Z^\mu(Q)$ is the weak current of the nucleus, $\lambda = -G_F m_Z^2 Q^2 / [2\sqrt{2}\pi(m_Z^2 - Q^2)]$, $G_F = g^2/4\sqrt{2}m_Z^2 \cos^2\theta_W$, $g$ is the weak coupling constant associated with $SU(2)$ and $\theta_W$ is the Weinberg angle. In the Weinberg-Salam model $g_V = -1/2 + 2x_W$, $g_A = -1/2$ and $x_W \equiv \sin^2\theta_W$.

The scattering cross section of the process is determined by

$$\sigma = \frac{4m_e^2 \varepsilon'}{f\varepsilon} \overline{\sum_{if}} |M_{fi}|^2 = \frac{\varepsilon'}{4f\varepsilon Q^4} R, \quad (2)$$

where $f$ is the recoil factor [3]. The notation $\overline{\sum_{if}}$ implies the average over the initial spin states and the summation over the final spin states, which are performed for both lepton ($L$) and nucleus ($H$), and we call briefly the summation. After the summation, we obtain the factor $R$ as follows [6,7]:

$$R = R_F + R_{FZ} + R_Z, \quad (3a)$$

$$R_F = \alpha^2 L_{\mu\nu} H_F^{\mu\nu}, \qquad R_{FZ} = 2\alpha\lambda L_{\mu\nu}^1 H_{FZ}^{\mu\nu}, \qquad R_Z = \lambda^2 L_{\mu\nu}^2 H_Z^{\mu\nu}, \quad (3b)$$

$$H_F^{\mu\nu} = \overline{\sum_{if(H)}} J_F^{\mu*} J_F^\nu, \qquad H_{FZ}^{\mu\nu} = \frac{1}{2}\overline{\sum_{if(H)}}(J_F^{\mu*} J_Z^\nu + J_Z^{\mu*} J_F^\nu), \qquad H_Z^{\mu\nu} = \overline{\sum_{if(H)}} J_Z^{\mu*} J_Z^\nu. \quad (4a\text{-}c)$$

The quantities $L_{\mu\nu}, L_{\mu\nu}^1, L_{\mu\nu}^2$ are lepton tensors and $H_F^{\mu\nu}, H_{FZ}^{\mu\nu}, H_Z^{\mu\nu}$ are nuclear tensors, which are also called the hadron tensors when considering the hadrons instead of nuclei. We see that the quantities $R_F$, $R_{FZ}$ and $R_Z$ are the contraction products of a lepton tensor and a nuclear tensor.

The high-energy electrons have almost only longitudinal polarizations, so their polarization vectors have two values $\xi = \pm 1$. If $S^\mu$ is the 4-dimensional vector of electron polarization, then it is possible to express approximately through the electron momentum by $S^\mu \approx \xi K^\mu/m_e$ and $S'^\mu \approx \xi' K'^\mu/m_e$. After performing the summation over lepton states and taking the matrix traces, we obtain

$$L_{\mu\nu} = 16m_e^2 \overline{\sum_{if(L)}} (\bar{u}'\gamma_\mu u)^*(\bar{u}'\gamma_\nu u) = 4[(1 + \xi\xi')X_{\mu\nu} + (\xi + \xi')Y_{\mu\nu}], \quad (5a)$$

$$X_{\mu\nu} = K_\mu K'_\nu + K'_\mu K_\nu + g_{\mu\nu} Q^2/2, \qquad Y_{\mu\nu} = -i\varepsilon_{\mu\nu\alpha\beta} K^\alpha K'^\beta, \, (\varepsilon_{0123} = 1), \quad (5b)$$

$$L_{\mu\nu}^1 = 16m_e^2 \overline{\sum_{if(L)}} (\bar{u}'\gamma_\mu u)^*(\bar{u}'\gamma_\nu(g_V + g_A\gamma_5)u) = g_V L_{\mu\nu} + g_A L'_{\mu\nu}, \quad (5c)$$

$$L'_{\mu\nu} = 4[(\xi + \xi')X_{\mu\nu} + (1 + \xi\xi')Y_{\mu\nu}], \quad (5d)$$

$$L_{\mu\nu}^2 = 16m_e^2 \overline{\sum_{if(L)}} (\bar{u}'\gamma_\mu(g_V + g_A\gamma_5)u)^*(\bar{u}'\gamma_\nu(g_V + g_A\gamma_5)u) = (g_V^2 + g_A^2)L_{\mu\nu} + 2g_V g_A L'_{\mu\nu}. \quad (5e)$$

From (5b) we see that $X_{\mu\nu} = X_{\nu\mu}$ and $Y_{\mu\nu} = -Y_{\nu\mu}$. The above formulas are calculated for lepton (electron) tensors. The summation of nuclear states will be performed in the next step.

We use a Cartesian coordinate system with the following unit vectors: $\mathbf{e}_z = \underline{\mathbf{q}}$, $\mathbf{e}_y = \underline{\mathbf{k} \times \mathbf{k}'}$, $\mathbf{e}_x = \underline{(\mathbf{k} \times \mathbf{k}') \times \mathbf{q}}$, where $\underline{\mathbf{a}} \equiv \mathbf{a}/|\mathbf{a}|$, i.e. the $OZ$ axis is along the direction of the momentum transfer $\mathbf{q}$, the $OX$ axis lies on the scattering plane, and the $OY$ axis is perpendicular to this plane. The next step is to change to the cyclic coordinate system $\zeta_0 = \mathbf{e}_z$, $\zeta_{\pm 1} = -(\pm\mathbf{e}_x + i\mathbf{e}_y)/\sqrt{2}$, and we shall expand the electromagnetic and weak transition currents of the nucleus into multipoles in this system

$$\rho(\mathbf{q}) = \sum_{Lm} \sqrt{4\pi(2L+1)} D_{0m}^{L*}(\gamma, \beta, 0) S_{Lm}^C(q), \quad (6a)$$

$$\mathbf{J}(\mathbf{q}) = \sum_{Lmp} \sqrt{4\pi(2L+1)} D_{pm}^{L*}(\gamma, \beta, 0) S_{Lm}^p(q)\zeta_p^*, \quad (p = 0, \pm 1). \quad (6b)$$

The coefficient $S_{Lm}^C$ is the Coulomb component of the expansion, three other coefficients $S_{Lm}^p$ are composed of the longitudinal component $S_{Lm}^0 \equiv S_{Lm}^\parallel$, and two transverse components which can be written as $S_{Lm}^{\pm 1} \equiv -(S_{Lm}^E \pm S_{Lm}^M)/\sqrt{2}$, where $S_{Lm}^E$ is the electric component and $S_{Lm}^M$ is the magnetic one. The inverse expressions of (6) are

$$S_{Lm}^C(q) = i^L \int \rho(\mathbf{r}) A_{Lm}^C(q, \mathbf{r}) d^3\mathbf{r}, \qquad S_{Lm}^\parallel(q) = i^{L-1} \int \mathbf{J}(\mathbf{r}) \cdot \mathbf{A}_{Lm}^\parallel(q, \mathbf{r}) d^3\mathbf{r}, \quad (7a,b)$$

$$S_{Lm}^E(q) = i^{L+1} \int \mathbf{J}(\mathbf{r}) \cdot \mathbf{A}_{Lm}^E(q, \mathbf{r}) d^3\mathbf{r}, \qquad S_{Lm}^M(q) = i^L \int \mathbf{J}(\mathbf{r}) \cdot \mathbf{A}_{Lm}^M(q, \mathbf{r}) d^3\mathbf{r}, \quad (7c,d)$$

where $A_{Lm}^C(q, \mathbf{r}), \mathbf{A}_{Lm}^\parallel(q, \mathbf{r}), \mathbf{A}_{Lm}^E(q, \mathbf{r})$ and $\mathbf{A}_{Lm}^M(q, \mathbf{r})$ are the basic multipole fields [8], of the Coulomb, longitudinal, electric and magnetic types, respectively. We rewrite here the general form of the structure of the electromagnetic and weak currents in the unified theory [4]:

$$J_\mu^F = V_\mu^{00} + V_\mu^{10}, \quad J_\mu^Z = V_\mu + A_\mu: \quad V_\mu = \beta_V^{(0)} V_\mu^{00} + \beta_V^{(1)} V_\mu^{10}, \quad A_\mu = \beta_A^{(0)} A_\mu^{00} + \beta_A^{(1)} A_\mu^{10}, \tag{8}$$

where $V_\mu$ is the vector weak current and $A_\mu$ is the pseudovector weak current, which is also called the axial current. Two superscripts of $V_\mu$ and $A_\mu$ in (8) express the isospin components (0 and 1) together with their principal projections. The expansions (6) and (7) are performed for three currents: when $J_\mu = J_\mu^F$ then $S_{Lm}^X = F_{Lm}^X$, and when $J_\mu = J_\mu^Z$ then $S_{Lm}^X = Z_{Lm}^X$ $(Z = V + A)$.

For computing the scattering cross section, we need to compute first the transition amplitude of the current (6), or equivalently, of their multipole components (7), between the initial and final states. On the other hand, $\hat{S}_{Lm}^X$ is a spherical tensor, so according to the Wigner-Eckart theorem we have $\langle J'M'|\hat{S}_{Lm}^X|JM\rangle = (-1)^{2L} C_{JMLm}^{J'M'} \langle J'||\hat{S}_L^X||J\rangle / \sqrt{2J'+1}$. The factor $\langle J'||\hat{S}_L^X||J\rangle$ is the reduced matrix element and will be denoted by $S_L^X$, with $S = F, V, A$ and $X = C, ||, E, M$. The quantities $F_L^X, V_L^X, A_L^X$ are also called the electromagnetic, vector and axial form factors, respectively. Since the currents $J_\mu^F$ and $V_\mu$ are conserved, the longitudinal component is expressed in terms of the Coulomb component as $F_{Lm}^{||} = \omega F_{Lm}^C / q$, $V_{Lm}^{||} = \omega V_{Lm}^C / q$. The axial current is not conserved, so there is no similar relation for it. Note that the energy transfer is the nuclear excitation energy and it can be neglected in comparison with the scattering energies of the order GeV. Consequently, we shall neglect longitudinal quantities when performing calculations and at the same time we have $\varepsilon' \approx \varepsilon$.

Now we express the scattering cross section through major physical quantities, firstly the electron polarizations, then the electron tensors in the expression (5). Three terms of the cross section (3) can be written in the form

$$R_F = 4\alpha^2[(1+\xi\xi')A_1 + (\xi+\xi')A_2], \tag{9}$$

$$R_{FZ} = 8\alpha\lambda\left\{[g_V(1+\xi\xi') + g_A(\xi+\xi')]B_1 + [g_V(\xi+\xi') + g_A(1+\xi\xi')]B_2\right\}, \tag{10}$$

$$R_Z = 4\lambda^2\left\{[(g_V^2+g_A^2)(1+\xi\xi') + 2g_Vg_A(\xi+\xi')]C_1 + [(g_V^2+g_A^2)(\xi+\xi') + 2g_Vg_A(1+\xi\xi')]C_2\right\}, \tag{11}$$

where $A_1 = X_{\mu\nu}H_F^{\mu\nu}$, $A_2 = Y_{\mu\nu}H_F^{\mu\nu}$, $B_1 = X_{\mu\nu}H_{FZ}^{\mu\nu}$, $B_2 = Y_{\mu\nu}H_{FZ}^{\mu\nu}$, $C_1 = X_{\mu\nu}H_Z^{\mu\nu}$ and $C_2 = Y_{\mu\nu}H_Z^{\mu\nu}$.

We shall compute the expressions $A_1, \ldots, C_2$, where longitudinal quantities are eliminated. The kinematic coefficients $u_C = 2\varepsilon\varepsilon' + Q^2/2 \approx 2\varepsilon^2(1-x^2)$, $u_T = k_t^2 - Q^2/2 \approx 2\varepsilon^2(1+x^2)$ and $u_T' = \varepsilon' k_\parallel - \varepsilon k_\parallel' \approx 2\varepsilon^2 x$ appear when computing the contraction products, where $x \equiv \sin(\theta/2)$ and $\theta$ is the scattering angle. First, we write the coefficient $A_1$ explicitly as

$$A_1 = \sum_{if(H)} \left\{(2\varepsilon\varepsilon' + \frac{Q^2}{2})\rho_F^*\rho_F - \frac{Q^2}{2}\mathbf{J}_F^*\cdot\mathbf{J}_F + 2\operatorname{Re}(\mathbf{k}\cdot\mathbf{J}_F^*)(\mathbf{k}'\cdot\mathbf{J}_F) - 2\varepsilon\operatorname{Re}\rho_F^*(\mathbf{k}'\cdot\mathbf{J}_F) - 2\varepsilon'\operatorname{Re}\rho_F^*(\mathbf{k}\cdot\mathbf{J}_F)\right\}. \tag{12}$$

Now we expand the expression (12) into the multipole form factors with the help of (6). The two last terms are equal to zero since their expressions contain the Clebsch-Gordan coefficients equal to zero when expanding the transition currents. The first three terms give

$$A_1 = H\sum_L \left\{u_C(F_L^C)^2 + u_T[(F_L^E)^2 + (F_L^M)^2]\right\}. \tag{13}$$

Next, by expanding $A_2$ in the same way, we get

$$A_2 = -i(\varepsilon\mathbf{k}' - \varepsilon'\mathbf{k})\cdot\overline{\sum_{if(H)}}(\mathbf{J}_F^* \times \mathbf{J}_F) - i(\mathbf{k}\times\mathbf{k}')\cdot\overline{\sum_{if(H)}}(\rho_F^*\mathbf{J}_F - \rho_F\mathbf{J}_F^*) = 0. \tag{14}$$

The first term is zero because its expression is the sum of the products $F_L^E F_L^M$ that violate the parity rule. The remaining term is also zero because the resulting expansions for $\rho_F^*\mathbf{J}_F$ and $\rho_F\mathbf{J}_F^*$ are both vectors with only longitudinal components, while $\mathbf{k}\times\mathbf{k}'$ is a vector with only transverse components.

Similarly, we also have

$$B_1 = \overline{\sum_{if(H)}}\left\{(2\varepsilon\varepsilon' + \frac{Q^2}{2})\operatorname{Re}(\rho_F^*\rho_Z) - \frac{Q^2}{2}\operatorname{Re}(\mathbf{J}_F^*\cdot\mathbf{J}_Z) + 2\operatorname{Re}(\mathbf{k}\cdot\mathbf{J}_F^*)(\mathbf{k}'\cdot\mathbf{J}_Z) - 2\varepsilon\operatorname{Re}\rho_F^*(\mathbf{k}'\cdot\mathbf{J}_Z) - 2\varepsilon'\operatorname{Re}\rho_Z^*(\mathbf{k}\cdot\mathbf{J}_F)\right\}.$$

The last two terms are zero for the same reasons as in $A_1$, while the other terms give

$$B_1 = H\sum_L[u_C F_L^C V_L^C + u_T(F_L^E V_L^E + F_L^M V_L^M)]. \tag{15}$$

The coefficient $B_2$ in the cyclic coordinate system has the same form as (14), but it corresponds to the products of two different currents. The resulting expansion of the second term is also zero due to a similar reason and the first term gives

$$B_2 = Hu_T'\sum_L(F_L^E A_L^M + F_L^M A_L^E). \tag{16}$$

Finally, multipole expansions for the coefficients $C_1$ and $C_2$ are also performed for each term in turn as for $B_1$ and $B_2$. The results are different from the previous ones, but the steps of calculation are quite similar. In addition, there are also zero terms like in (15) and (16). The none-zero remaining terms give the following:

$$C_1 = H\sum_L\left\{u_C[(V_L^C)^2 + (A_L^C)^2] + u_T[(V_L^E)^2 + (V_L^M)^2 + (A_L^E)^2 + (A_L^M)^2]\right\}, \tag{17}$$

$$C_2 = 2Hu_T'\sum_L(V_L^E A_L^M + V_L^M A_L^E). \tag{18}$$

The expressions (9)-(11), with the coefficients determined by (13)-(18), are the complete multipole expansion for the high-energy scattering cross section of polarized electrons from unoriented nuclei. Our multipole expansion for $R_F$ essentially coincides with the results of Weigert-Rose [1], but there are some changes for being compatible with high energies [6,7,9].

## 3. Construction of multipole operators

We shall determine the expressions of multipole operators within the multiparticle shell model. Using $V_\mu^{00} = \bar{\psi}\gamma_\mu\psi/2$, $V_\mu^{10} = \bar{\psi}\gamma_\mu\tau_3\psi/2$, $A_\mu^{00} = \bar{\psi}\gamma_\mu\gamma_5\psi/2$, $A_\mu^{10} = \bar{\psi}\gamma_\mu\gamma_5\tau_3\psi/2$, where the nucleon wavefunction is a Dirac spinor with the components $\phi = \boldsymbol{\sigma}.\mathbf{p}\chi/(\varepsilon_N - m_N)$ and $\chi = \boldsymbol{\sigma}.\mathbf{p}\phi/(\varepsilon_N + m_N)$, we first deduce the nucleon currents in the nucleus according to (8) as follows

$$\rho_F(\mathbf{r}) = (\varepsilon^s + \varepsilon^v\tau_3)\delta(\mathbf{r}), \qquad \mathbf{J}_F(\mathbf{r}) = \mathbf{J}_e(\mathbf{r}) + \mathbf{J}_m(\mathbf{r}), \qquad \mathbf{J}_m(\mathbf{r}) = \nabla \times \boldsymbol{\mu}(\mathbf{r}), \tag{19a}$$

$$\mathbf{J}_e(\mathbf{r}) = \frac{1}{m_N}(\varepsilon^s + \varepsilon^v\tau_3)\delta(\mathbf{r})\mathbf{p}, \qquad \boldsymbol{\mu}(\mathbf{r}) = \frac{1}{2m_N}(\gamma^s + \gamma^v\tau_3)\delta(\mathbf{r})\boldsymbol{\sigma}, \tag{19b}$$

$$\rho_A(\mathbf{r}) = \frac{1}{m_N}(\beta_A^{(0)}\eta^s + \beta_A^{(1)}\eta^v\tau_3)\delta(\mathbf{r})(\boldsymbol{\sigma}.\mathbf{p}), \qquad \mathbf{J}_A(\mathbf{r}) = (\beta_A^{(0)}\eta^s + \beta_A^{(1)}\eta^v\tau_3)\delta(\mathbf{r})\boldsymbol{\sigma}. \tag{19c}$$

The vector current is proportional to the electromagnetic current through the parameters of the electroweak theory. The components of nucleon currents are $\varepsilon^s = (\varepsilon_p + \varepsilon_n)/2$, $\varepsilon^v = (\varepsilon_p - \varepsilon_n)/2$, $\gamma^s = (\gamma_p + \gamma_n)/2$, $\gamma^v = (\gamma_p - \gamma_n)/2$, $\eta^s = (\eta_p + \eta_n)/2$ and $\eta^v = (\eta_p - \eta_n)/2$. We use $\varepsilon_p, \varepsilon_n, \gamma_p, \gamma_n, \eta_p$ and $\eta_n$ as parameters indicating the electromagnetic characteristics of nucleon when regarding it as a point-like. When taking into account its size, they will be converted to the nucleon form factors. Meanwhile, the isospin components become $G_E^s = (G_E^p + G_E^n)/2$, $G_E^v = (G_E^p - G_E^n)/2$, $G_M^s = (G_M^p + G_M^n)/2$, $G_M^v = (G_M^p - G_M^n)/2$, $G_A^s = (G_A^p + G_A^n)/2 \equiv G_P$ and $G_A^v = (G_A^p - G_A^n)/2 \equiv G_A$. At high energies, all nucleon form factors are parameterized in terms of dipoles [10–12].

The nuclear currents regarded as the sums of nucleon currents are determined by

$$\rho_F(\mathbf{r}) = \sum_a X_a^F \delta(\mathbf{r} - \mathbf{r}_a), \qquad \mathbf{J}_e(\mathbf{r}) = \frac{1}{2m_N}\sum_a X_a^F \{\delta(\mathbf{r} - \mathbf{r}_a)\mathbf{p}_a\}_{sym}, \tag{20a}$$

$$\mathbf{J}_m(\mathbf{r}) = \nabla \times \boldsymbol{\mu}(\mathbf{r}), \qquad \boldsymbol{\mu}(\mathbf{r}) = \frac{1}{2m_N}\sum_a Y_a^F \delta(\mathbf{r} - \mathbf{r}_a)\boldsymbol{\sigma}_a, \tag{20b}$$

$$\rho_A(\mathbf{r}) = \frac{1}{m_N}\sum_a Y_a^A \delta(\mathbf{r} - \mathbf{r}_a)\boldsymbol{\sigma}_a.\mathbf{p}_a, \qquad \mathbf{J}_A(\mathbf{r}) = \sum_a Y_a^A \delta(\mathbf{r} - \mathbf{r}_a)\boldsymbol{\sigma}_a. \tag{20c}$$

The nucleon momentum is symmetrized due to the conserved current. Here, we introduce the notations $X_a^F = (G_E^s + G_E^v\tau_{a3})$, $Y_a^F = (G_M^s + G_M^v\tau_{a3})$, $X_a^V = (\beta_V^{(0)}G_E^s + \beta_V^{(1)}G_E^v\tau_{a3})$, $Y_a^V = (\beta_V^{(0)}G_M^s + \beta_V^{(1)}G_M^v\tau_{a3})$ and $Y_a^A = (\beta_A^{(0)}G_A^s + \beta_A^{(1)}G_A^v\tau_{a3})$, where $a$ is the particle index. The vector current can be inferred from the electromagnetic current by replacing $X_a^F \to X_a^V$ and $Y_a^F \to Y_a^V$. The expressions of pseudovector current (20c) derived from Dirac matrices allow us to compute the axial form factors directly without impulse approximation. From (7) and (20), we arrive at the detailed expressions of multipole operators as follows

$$\hat{F}_{Lm}^C(q) = i^L \sum_a X_a^F j_L(qr_a)Y_{Lm}(\underline{\mathbf{r}}_a), \qquad \hat{F}_{Lm}^M(q) = \frac{i^{L+1}q}{2m_N[L]}\sum_a [2X_a^F \mathbf{A}_a^L.\mathbf{l}_a + Y_a^F \mathbf{B}_a^L.\boldsymbol{\sigma}_a], \tag{21a,b}$$

$$\hat{F}_{Lm}^E(q) = \frac{i^L q}{2m_N}\sum_a \left\{ \frac{X_a^F}{\sqrt{L(L+1)}}[r_a j_L(qr_a)d_a + d_a r_a j_L(qr_a)]Y_{Lm}(\underline{\mathbf{r}}_a) + Y_a^F j_L(qr_a)\mathbf{Y}_{Lm}^L(\underline{\mathbf{r}}_a).\boldsymbol{\sigma}_a \right\}, \tag{21c}$$

$$\hat{A}_{Lm}^C(q) = -\frac{i^{L+1}}{m_N}\sum_a Y_a^A j_L(qr_a)Y_{Lm}(\underline{\mathbf{r}}_a)\nabla_a.\boldsymbol{\sigma}_a, \tag{22a}$$

$$\hat{A}_{Lm}^E(q) = \frac{i^{L+1}}{[L]}\sum_a Y_a^A \mathbf{B}_a^L.\boldsymbol{\sigma}_a, \qquad \hat{A}_{Lm}^M(q) = i^L \sum_a Y_a^A j_L(qr_a)\mathbf{Y}_{Lm}^L(\underline{\mathbf{r}}_a).\boldsymbol{\sigma}_a. \tag{22b,c}$$

The notations $[L] \equiv \sqrt{2L+1}$, $d_a \equiv d/dr_a$, $\mathbf{A}_a^L \equiv \mathbf{A}^L(\mathbf{r}_a)$ and $\mathbf{B}_a^L \equiv \mathbf{B}^L(\mathbf{r}_a)$ are used for brevity, where

$$\mathbf{A}^L(\mathbf{r}) = \frac{1}{\sqrt{L+1}}j_{L-1}(qr)\mathbf{Y}_{Lm}^{L-1}(\underline{\mathbf{r}}) + \frac{1}{\sqrt{L}}j_{L+1}(qr)\mathbf{Y}_{Lm}^{L+1}(\underline{\mathbf{r}}), \tag{23a}$$

$$\mathbf{B}^L(\mathbf{r}) = \sqrt{L+1}j_{L-1}(qr)\mathbf{Y}_{Lm}^{L-1}(\underline{\mathbf{r}}) - \sqrt{L}j_{L+1}(qr)\mathbf{Y}_{Lm}^{L+1}(\underline{\mathbf{r}}). \tag{23b}$$

The condition for the validity of the non-relativistic treatment of nuclei requires that the Darwin term [2] in the Coulomb matrix elements needs to be discarded at high energies. To determine multipole form factors, it is necessary to compute the reduced matrix elements of (21) and (22).

## 4. General formulas of matrix elements

Now we use the one-particle fractional parentage coefficients [8,13,14] to infer some general formulas of reduced matrix elements since all multipole operators depend only on one-particle variables. We denote $|i\rangle \equiv |n_i L_i S_i T_i M_{T_i} J_i\rangle$ and $|f\rangle \equiv |n_f L_f S_f T_f M_{T_f} J_f\rangle$ as the initial and final states in $LS$ coupling and curly brackets $\{...\}$ as the $6j$ or $9j$ coefficients. The total wavefunction is written as a product of the orbital, spin and isospin wavefunctions. Furthermore, nucleons are considered to be identical so they have the same quantum numbers in the same shell. From the formulas for calculating the reduced matrix elements of tensor operators [15,16], it can be proved that

$$\langle f \| \sum_a X_a^F \Im_L(a) \| i \rangle = \delta_{S_i S_f} \delta_{M_{T_i} M_{T_f}} A' \sum_P (-1)^{L_P + S_i + J_i + l_i} [J_i][J_f][L_i][L_f] \langle \psi_f\{|\psi_P\rangle \langle \psi_i\{|\psi_P\rangle$$

$$\times \begin{Bmatrix} L_i & J_i & S_i \\ J_f & L_f & L \end{Bmatrix} \begin{Bmatrix} l_i & L_i & L_P \\ L_f & l_f & L \end{Bmatrix} \langle n_f l_f \| \Im_L \| n_i l_i \rangle Q_P^F, \quad (24a)$$

$$\langle f \| \sum_a Y_a^F j_{L'}(qr_a) \mathbf{Y}_L^{L'}(\mathbf{r}_a) \cdot \boldsymbol{\sigma}_a \| i \rangle = \delta_{M_{T_i} M_{T_f}} \sqrt{6} A' \sum_P (-1)^{L_P + S_P + L_f + S_f + l_i + L' + 3/2} [L][J_i][J_f][L_i][L_f]$$

$$\times [S_i][S_f] \langle \psi_f\{|\psi_P\rangle \langle \psi_i\{|\psi_P\rangle \begin{Bmatrix} L_f & S_f & J_f \\ L_i & S_i & J_i \\ L' & 1 & L \end{Bmatrix} \begin{Bmatrix} l_i & L_i & L_P \\ L_f & l_f & L' \end{Bmatrix} \begin{Bmatrix} 1/2 & S_i & S_P \\ S_f & 1/2 & 1 \end{Bmatrix}$$

$$\times \langle n_f l_f \| j_{L'}(qr) \mathbf{Y}^{L'}(\mathbf{r}) \| n_i l_i \rangle Q_P^F, \quad (24b)$$

$$\langle f \| \sum_a X_a^F j_{L'}(qr_a) \mathbf{Y}_L^{L'}(\mathbf{r}_a) \cdot \mathbf{l}_a \| i \rangle = \delta_{S_i S_f} \delta_{M_{T_i} M_{T_f}} A' \sum_P (-1)^{L_P + S_i + J_i + l_f + L} [L][J_i][J_f][L_i][L_f]$$

$$\times \langle \psi_f\{|\psi_P\rangle \langle \psi_i\{|\psi_P\rangle \begin{Bmatrix} L_i & J_i & S_i \\ J_f & L_f & L \end{Bmatrix} \begin{Bmatrix} l_i & L_i & L_P \\ L_f & l_f & L \end{Bmatrix} \begin{Bmatrix} L' & 1 & L \\ l_i & l_f & l_i \end{Bmatrix}$$

$$\times \langle n_f l_f \| j_{L'}(qr) \mathbf{Y}^{L'}(\mathbf{r}) \| n_i l_i \rangle \langle l_i \| \mathbf{l} \| l_i \rangle Q_P^F, \quad (24c)$$

$$\langle i \| \sum_a Y_a^A j_{L'}(qr_a) \mathbf{Y}_L^{L'}(\mathbf{r}_a) \cdot \boldsymbol{\sigma}_a \| f \rangle = \delta_{M_{T_i} M_{T_f}} \sqrt{6} A' \sum_P (-1)^{L_P + S_P + L_i + S_i + l_f + L' + 3/2} [L][J_i][J_f][L_i][L_f]$$

$$\times [S_i][S_f] \langle \psi_f\{|\psi_P\rangle \langle \psi_i\{|\psi_P\rangle \begin{Bmatrix} L_i & S_i & J_i \\ L_f & S_f & J_f \\ L' & 1 & L \end{Bmatrix} \begin{Bmatrix} l_f & L_f & L_P \\ L_i & l_i & L' \end{Bmatrix} \begin{Bmatrix} 1/2 & S_f & S_P \\ S_i & 1/2 & 1 \end{Bmatrix}$$

$$\times \langle n_i l_i \| j_{L'}(qr) \mathbf{Y}^{L'}(\mathbf{r}) \| n_f l_f \rangle Q_P^A, \quad (24d)$$

$$Q_P^F = \delta_{T_i T_f} G_{E,M}^s + (-1)^{T_P + T_f + 3/2} \sqrt{6} [T_i] C_{T_i M_{T_i} 10}^{T_f M_{T_f}} \begin{Bmatrix} 1/2 & T_i & T_P \\ T_f & 1/2 & 1 \end{Bmatrix} G_{E,M}^v, \quad (25a)$$

$$Q_P^A = \delta_{T_i T_f} \beta_A^{(0)} G_A^s + (-1)^{T_P + T_f + 3/2} \sqrt{6} [T_i] C_{T_i M_{T_i} 10}^{T_f M_{T_f}} \begin{Bmatrix} 1/2 & T_i & T_P \\ T_f & 1/2 & 1 \end{Bmatrix} \beta_A^{(1)} G_A^v. \quad (25b)$$

The coefficient $A'$ is the nucleon number in the state $nl^{A'}$ and $\Im_L$ stands for the operators acting only on the orbital wavefunction. The isospin matrix elements $Q_P = Q_0, Q_1$ correspond to $T_P = 0, 1$. The summations are taken over the parent states $P = L_P, S_P, T_P$ and $\langle \psi\{|\psi_P\rangle$ is the fractional parentage coefficient. The electromagnetic form factors are calculated completely thanks to the above formulas. Also, the vector form factors can be determined by replacing $X_a^F, Y_a^F \to X_a^V, Y_a^V$ and

$$Q_P^F \to Q_P^V = \delta_{T_i T_f} \beta_V^{(0)} G_{E,M}^s + (-1)^{T_P + T_f + 3/2} \sqrt{6} [T_i] C_{T_i M_{T_i} 10}^{T_f M_{T_f}} \begin{Bmatrix} 1/2 & T_i & T_P \\ T_f & 1/2 & 1 \end{Bmatrix} \beta_V^{(1)} G_{E,M}^v. \quad (26)$$

In addition, the Coulomb axial form factors are determined by integrating by parts and then applying (24d). Finally, the other axial form factors are also computed through (24) by replacing $Y_a^F \to Y_a^A$ and $Q_P^F \to Q_P^A$. It can be seen that when the fractional parentage coefficients are set equal to unity, we can deduce the familiar formulas of reduced matrix elements for the case of one and two particles.

## 5. Results and discussions

In principle, the formulas given above allow us to determine the multipole form factors of light nuclei with arbitrary spins. In this paper, we select the spin-3/2 nucleus for illustration.

### 5.1. Elastic scattering $e^- - {}^7Li$ in ground state $(3/2 \to 3/2)$

For a detailed calculation, we consider three valence nucleons in the $1p$ shell. The parent states are $^{13}S$, $^{31}S$, $^{13}D$ and $^{31}D$ two-nucleon states, where states are denoted by $^{2T+1, 2S+1}L_P$. The initial and final states belong to the $^{22}P$ doublet so the products of two one-particle fractional parentage coefficients are replaced by the squares of a single coefficient. Their values can be easily found in the tables of Jahn and Van Wieringen [2,14]. The transition matrix elements are calculated for the states $n_i = n_f = 1$, $L_i = L_f = 1$, $l_i = l_f = 1$, $S_i = S_f = 1/2$, $J_i = J_f = 3/2$, $T_i = T_f = 1/2$ and $M_{T_i} = M_{T_f} = -1/2$. According to the selection rule, there are possible multipoles of $L = 0, 1, 2, 3$, where $\hat{F}_{Lm}^E$, $\hat{V}_{Lm}^E$ and $\hat{A}_{Lm}^M$ are absent. Noting that all multipole form factors are real, we arrive at

$$F_0^C(q) = \frac{1}{\sqrt{\pi}} \tilde{X}^F J_0, \qquad F_2^C(q) = \frac{3}{5\sqrt{\pi}} \tilde{X}^F J_2, \quad (27a,b)$$

$$F_1^M(q) = -\frac{\sqrt{5}q}{9\sqrt{2\pi} m_N} [\tilde{X}^F (J_0 + J_2) + 3\tilde{Y}^F (J_0 - \frac{3}{25} J_2)], \qquad F_3^M(q) = -\frac{3\sqrt{3}q}{5\sqrt{5\pi} m_N} \tilde{Y}^F J_2, \quad (27c,d)$$

$$A_1^C(q) = -\frac{\sqrt{5}q}{3\sqrt{\pi} m_N} \tilde{Y}^A (J_0 + \frac{6}{25} J_2), \qquad A_3^C(q) = -\frac{9q}{5\sqrt{5\pi} m_N} \tilde{Y}^A J_2, \quad (28a,b)$$

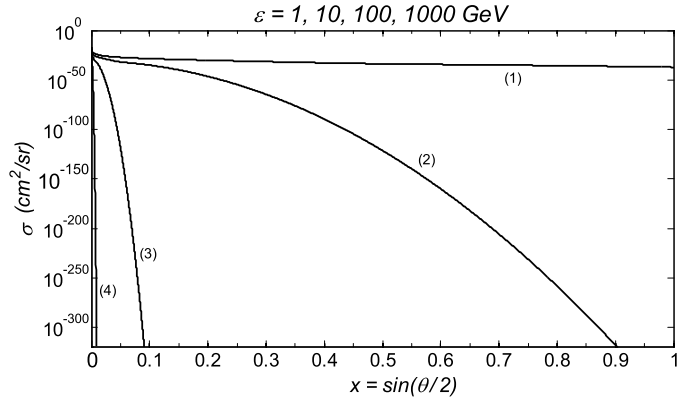

**Fig. 1.** Total cross section in $e^- - {}^7Li$ scattering at high energies.

$$A_1^E(q) = -\frac{\sqrt{10}}{3\sqrt{\pi}} \tilde{Y}^A (J_0 - \frac{3}{25} J_2), \qquad A_3^E(q) = -\frac{6\sqrt{3}}{5\sqrt{5\pi}} \tilde{Y}^A J_2, \qquad (28c,d)$$

where the notations $\tilde{X}^F = 3G_E^s - G_E^v$, $\tilde{Y}^F = G_M^s + G_M^v$ and $\tilde{Y}^A = \beta_A^{(0)} G_A^s + \beta_A^{(1)} G_A^v$ are used. The radial integrals $J_0(q)$ and $J_2(q)$ have the same expressions as given in [2] when using the harmonic oscillator wavefunctions. We do not write the vector form factors because they can be deduced from (27) by replacing $\tilde{X}^F \to \tilde{X}^V = 3\beta_V^{(0)} G_E^s - \beta_V^{(1)} G_E^v$ and $\tilde{Y}^F \to \tilde{Y}^V = \beta_V^{(0)} G_M^s + \beta_V^{(1)} G_M^v$.

### 5.2. Quasi-elastic scattering $e^- - {}^7Li$ in the transition $3/2 \to 1/2$

In this case, there are possible multipoles of $L = 1, 2$ and none of them are forbidden. Matrix elements are determined for the initial and final states similar to the elastic scattering, except that $J_i = 3/2$ and $J_f = 1/2$. These states also have the same parent states and the same fractional parentage coefficients as in the previous case. Applying the given formulas, we obtain

$$F_2^C(q) = -\frac{3}{5\sqrt{\pi}} \tilde{X}^F J_2, \qquad F_2^E(q) = \frac{3\sqrt{3}q}{10\sqrt{2\pi} m_N} \tilde{Y}^F J_2, \qquad (29a,b)$$

$$F_1^M(q) = \frac{q}{9\sqrt{2\pi} m_N}[\tilde{X}^F(J_0 + J_2) - 6\tilde{Y}^F(J_0 + \frac{3}{20} J_2)], \qquad (29c)$$

$$A_1^C(q) = \frac{2q}{3\sqrt{\pi} m_N} \tilde{Y}^A (J_0 - \frac{3}{10} J_2), \qquad A_1^E(q) = -\frac{2\sqrt{2}}{3\sqrt{\pi}} \tilde{Y}^A (J_0 + \frac{3}{20} J_2), \qquad A_2^M(q) = \frac{6\sqrt{3}}{10\sqrt{2\pi}} \tilde{Y}^A J_2. \qquad (30a-c)$$

Similarly, the vector form factors may also be inferred from the electromagnetic ones. In both cases, all the nuclear form factors depend on the nucleon form factors and the parameters of the electroweak theory. Also, they are proportional to $J_0(q)$ and $J_2(q)$ in different ways. Here, we use no approximation and do not omit the convection currents as well. If the axial form factors are determined by the relations proportional to the isospin components of the electromagnetic form factors as performed in [5], they will differ from the above results by the factor $q/2m_N$. This shows the difference between the directly calculated results and those by impulse approximation.

Now we analyze the numerical calculation results of the cross section for the case of unpolarized electrons within the Weinberg-Salam model. We need to multiply cross section by two since it was averaged over the initial electron states. The nucleon form factors are $G_E^p = -G_E^n/1.91\eta = G_M^p/2.79 = -G_M^n/1.91 = G_D$, $G_D = 1/(1 + q^2/0.71 \text{ GeV}^2)^2$, $\eta = -Q^2/4m_N^2$, $G_A = G_A(0)/(1 + q^2/m_A^2)^2$, where $G_A(0) = 1.27$ and $m_A = 1.026$ GeV [11,12]. The radial integrals are $J_0 = (1 - 2z/3)e^{-z}$ and $J_2 = 2ze^{-z}/3$, where $z = q^2/4\beta$ [2]. The harmonic oscillator parameter is chosen as $\beta^{-1} = 4$ GeV$^{-2}$ to be compatible with the dimension of momentum.

The graphs of the quasi-elastic and elastic cross sections at the energies of MeV are distributed in all scattering angles and have the same order of magnitude as in previous studies [2,17]. The quasi-elastic cross section is several orders of magnitude smaller than the elastic one. Also, our result differs from that of Willey only when the scattering angle is very large. This is due to the nucleon form factors depending on the momentum transfer. When using the nucleon form factors as given in [2] and adding the Darwin term, graphs of them nearly coincide with each other.

At the energies of GeV, the quasi-elastic and elastic cross sections get closer together and transform in the same manner as depicted in Fig. 1. Curves (1) and (2) show that the scattering processes with small angles are preferred since the cross section decreases rapidly at large angles. When the energies are up to hundreds of GeV and more, there are only the forward scattering processes as described by curves (3) and (4). The quasi-elastic cross section at the energies of GeV differs very little from that of the elastic scattering since the transitions under consideration belong to the two lowest states of doublet and we had set $\varepsilon = \varepsilon'$.

The graphs in Fig. 1 represent the total cross section with different energies of the incident electrons. They describe the dependence of cross section on scattering angles but do not show the contribution of weak interaction to the general expression. With the calculated form factors, three terms of the cross section can also be investigated separately. From then, we can examine the ratio $(\sigma_{FZ} + \sigma_Z)/\sigma_F$, for example in the elastic scattering as shown in Fig. 2, to estimate the contribution of weak interaction in the unified electroweak interaction. For incident electrons with energies less than tens of GeV, the weak interaction is several orders of magnitude smaller than the electromagnetic one, i.e. the role of the weak interaction is negligible. Corresponding to the energies of 50 GeV, 100 GeV, 500 GeV and 1000 GeV, curves (1)-(4) show that the weak interaction increases rapidly in scattering angles and the highest ratio is approximately 8.5%. It now can be measured simultaneously with the electromagnetic interaction.

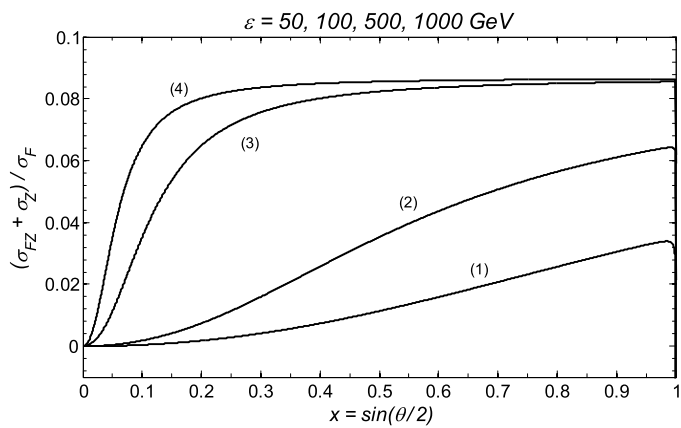

**Fig. 2.** The effect of weak interaction in $e^- - {}^7Li$ elastic scattering at high energies.

Moreover, curves also indicate that the ratio under consideration decreases rapidly at an angle of 180°. This comes from the fact that, at this scattering angle, the Coulomb component has a sharp minimum whereas the transverse component is nearly constant [18,19].

## 6. Conclusion

We have carried out the multipole expansion for the transition currents in the nucleus into the multipole form factors and expressed the cross section in terms of them. These factors are the reduced matrix elements of the multipole components which appear when expanding the transition currents. The multipole form factors are the simplest components of the transition currents, which can be computed directly from the nuclear structure models. Therefore, the method of multipole expansion allows us to obtain more detailed information about the nuclear structure, first of all, the information which manifests only in the high-energy processes. Also, we have used the multiparticle shell model and the fractional parentage coefficient method to give general formulas for the reduced matrix elements. As a result, all multipole form factors can be calculated directly without approximation. It also opens new perspectives in checking the unified electroweak interaction model. The computation of the multipole form factors for nuclei as well as the evaluation of the parameters of the electroweak theory will become an extensive area of the nuclear structure theory.

**Declaration of competing interest**

The authors declare that they have no known competing financial interests or personal relationships that could have appeared to influence the work reported in this paper.

**Data availability**

Data will be made available on request.

**Acknowledgements**

We gratefully acknowledge Professor K.A. Gridnev (Saint Petersburg State University, Russia) for helpful discussions while performing calculations and we were deeply saddened that he was deceased.

**Appendix A. Supplementary material**

Supplementary material related to this article can be found online at https://doi.org/10.1016/j.physletb.2023.138095.